\documentclass[twocolumn,aps,pre,superscriptaddress]{revtex4-2}

\usepackage{amsmath}
\usepackage{amssymb}
\usepackage{xcolor}
\usepackage{graphicx}
\usepackage{bm}
\usepackage{placeins}
\usepackage[colorlinks=true,linkcolor=blue,citecolor=blue]{hyperref}

\begin{document}

\title{Critical dynamics govern the evolution of political regimes}

\author{Joshua Uhlig}
\email{joshua.uhlig@uni-potsdam.de}
\affiliation{University of Potsdam, Institute for Physics \& Astronomy, 14476
Potsdam, Germany}
\author{Paula Pirker-D{\'i}az}
\affiliation{University of Potsdam, Institute for Physics \& Astronomy, 14476
Potsdam, Germany}
\author{Matthew Wilson}
\affiliation{University of South Carolina, Department of Political Science,
Columbia, SC, USA}
\author{Ralf Metzler}
\email{rmetzler@uni-potsdam.de}
\affiliation{University of Potsdam, Institute for Physics \& Astronomy, 14476
Potsdam, Germany}
\author{Karoline Wiesner}
\email{karoline.wiesner@uni-potsdam.de}
\affiliation{University of Potsdam, Institute for Physics \& Astronomy, 14476
Potsdam, Germany}

\date{\today}

\begin{abstract}
The emergence and decline of democratic systems worldwide raises fundamental
questions about the dynamics of political change. Contrary to the idea
of a stable endpoint of liberal democracy, recent backsliding towards
less democratic regimes highlights the non-stationary nature of regime
evolution. Here, we analyse the historical trajectories of countries within
a two-dimensional regime space derived from the principal components of the
Varieties of Democracy dataset. We observe weakly non-ergodic dynamics
unfolding in an effective landscape characterised by sparse and shifting
basins of stability. Step sizes and sojourn times characterising this
dynamics follow heavy-tailed distributions near the critical regime, in
which mean values appear to diverge. These facts point to the intermittent
and heterogeneous nature of the regime change dynamics. A continuous time
random walk model reproduces the dynamics of the three most recent decades
with remarkable accuracy. Together, these results suggest that some aspects
of political regime evolution follow universal stochastic principles, while
remaining punctuated by unique historical pathways.
\end{abstract}

\maketitle

\section{\label{sec:intro}Introduction}

Political systems evolve over time, but the dynamic nature of this change
remains poorly understood. Classical theories of modernisation and democratic
consolidation describe regime evolution as a gradual progression towards
stable governance \cite{Lipset1959,North1981,Fukuyama1989}. However, recent
episodes of democratic backsliding and regime reversal challenge this idea
\cite{BackslidingReview}. Researchers also argue that political development
is path-dependent, with each country following trajectories shaped by key
moments in its history \cite{Pierson2000, Mahoney2000}. It is not known
whether political regimes evolve along unique historical paths or whether
their trajectories are constrained by general dynamical laws.

\begin{figure}
\includegraphics[width=\columnwidth]{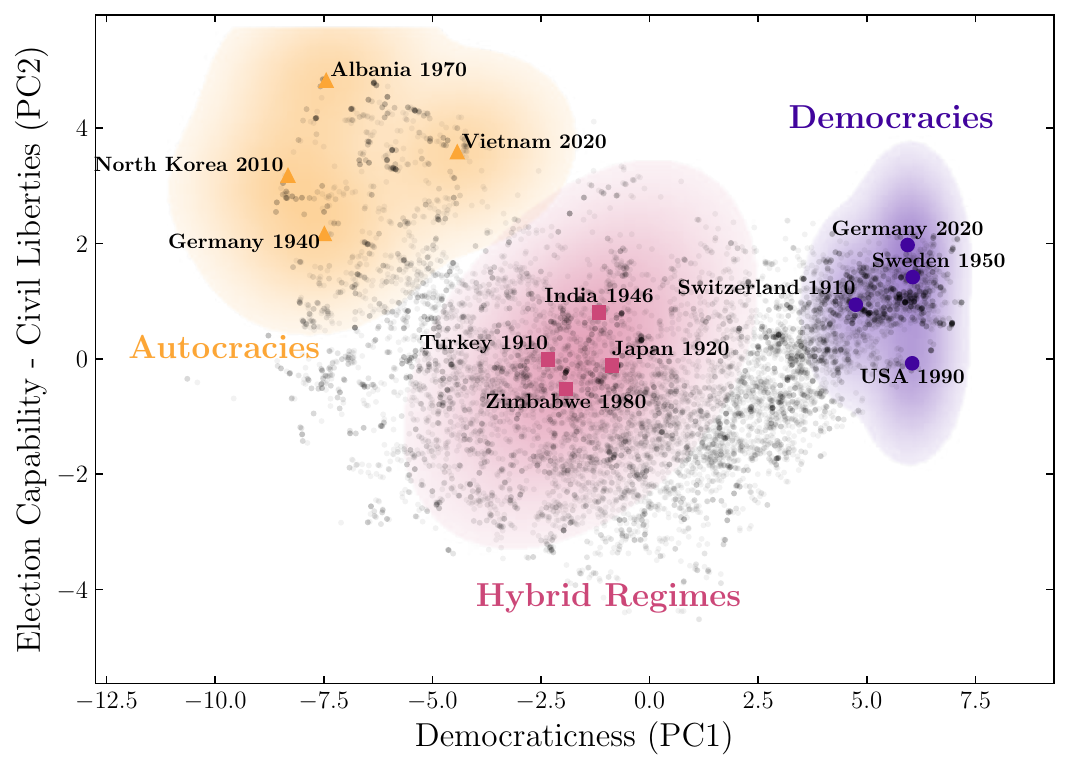}
\caption{\textbf{All country-year observations in the PC1-PC2 political
regime space.} Selected examples indicate regions typically associated with
democratic, autocratic, and hybrid regimes. These labels are illustrative
and do not imply discrete regime classifications.}
\label{fig:PCspace}
\end{figure}

To explore these dynamics, we draw on a subset of the Varieties of Democracy
(V-Dem) dataset \cite{VDem}, which provides yearly measures of political
characteristics for countries from 1900 to 2021. We represent each country's
position in a two-dimensional regime space defined by the first two principal
components (PC1 and PC2), following Ref.~\cite{WiesnerBienWilson2024}. PC1
corresponds to a democracy-autocracy axis, while PC2 reflects a trade-off
between electoral capabilities and civil liberties \cite{WiesnerBienWilson2024}.
The resulting two-dimensional PC1-PC2 plane is treated as a political regime
space. Figure~\ref{fig:PCspace} shows all country-year points in this regime
space, along with several well-known examples to provide intuition for its
structure.

As countries evolve, they move through this space along trajectories.
This suggests viewing regime evolution as a dynamical process
unfolding over time.  Approaches from sociophysics have increasingly
applied concepts from statistical physics to collective social phenomena
\cite{Castellano2009,Galam2012,Ubaldi2021}. Stochastic models developed in
statistical physics provide a natural framework for analysing such dynamics
\cite{Klages2008,Krapivsky2010}. Although originally developed to explain
the motion of particles in fluids \cite{Einstein1905,Langevin1908}, these
models have since been applied to animal movement \cite{Vilk2022,Meyer2023},
financial time series \cite{Cherstvy2017,Ritschel2021}, and human mobility
\cite{Brockmann2006}. Dedicated approaches to parameter regression,
classification of the stochastic processes underlying the observation, as
well as methods to identify change-points in the dynamics have been developed
recently \cite{N1,N2,N3,Feng2024}. In parallel, diffusion-based analyses
have been applied to political regime trajectories, uncovering dynamical
structure in regime stability \cite{PirkerDiaz2025}.

In our analysis, we find that regime trajectories exhibit near scale-free
step size and sojourn time distributions and display weak ergodicity
breaking. These features indicate that political regimes evolve through
intermittent, path-dependent dynamics. We show that these patterns can be
captured by a continuous time random walk (CTRW) \cite{MontrollWeiss1965}
parameterised from the empirical distributions. The model reproduces both
the average behaviour and the trajectory-level variability. Our findings
show that political regime change appears to be simultaneously historically
specific and dynamically universal.

\section{\label{sec:results}Results}

\subsection{Empirical regime dynamics}

\begin{figure}
\includegraphics[width=\columnwidth]{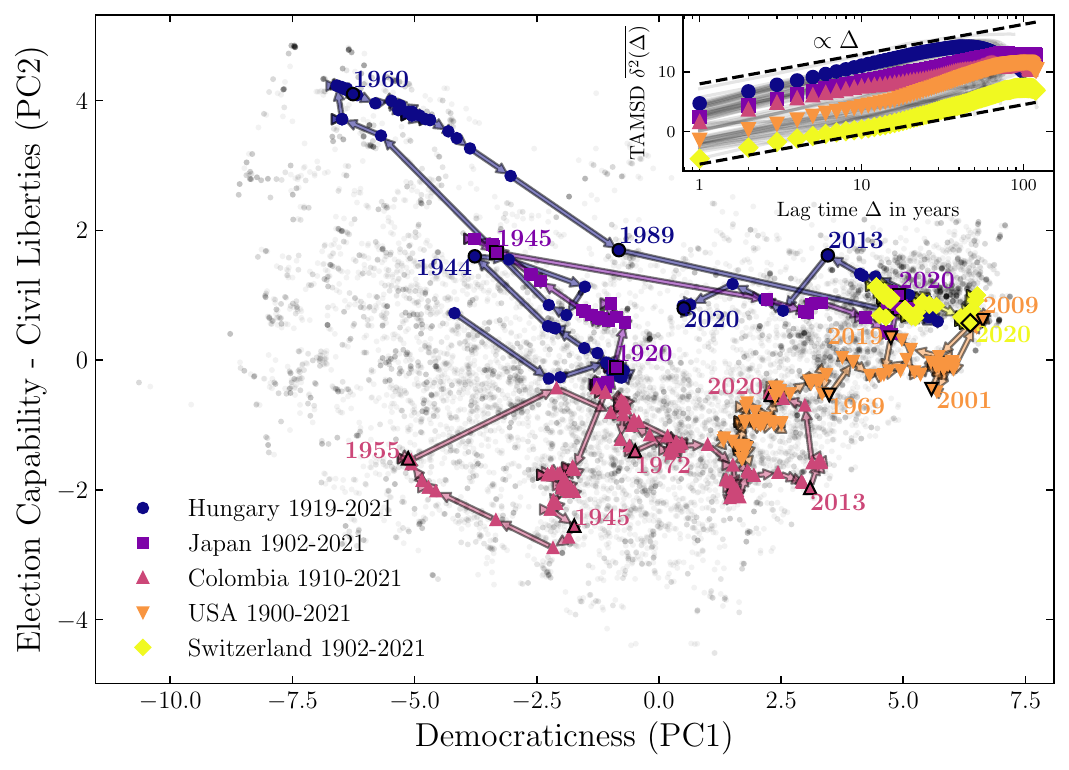}
\caption{\label{fig:fig1_trajectory}\textbf{Representative regime trajectories
in the two-dimensional political regime space.} Coloured symbols show selected
country trajectories over time, superimposed on all country-year observations
(black transparent dots). Inset: time-averaged mean-squared displacement
(TAMSD) for the same countries.}
\end{figure}

We analyse country trajectories in the two-dimensional regime space
defined by the first two principal components of the V-Dem dataset: PC1
(democraticness) and PC2 (trade-off between electoral capability and
civil liberties). Examples of some of the longest trajectories are shown
in Fig.~\ref{fig:fig1_trajectory}. They illustrate a broad spectrum of
dynamical behaviours, ranging from gradual shifts to abrupt transitions. Most
trajectories include periods with little or no change, sojourn times,
interspersed with sudden shifts whose magnitude and direction vary from year
to year.

Countries such as Hungary have explored large portions of the political space
over the past century, while Japan and Colombia remained confined to a few
regime regions separated by large jumps. Early in the 20th century, Japan
was a monarchy and occupied the hybrid regime region, then shifted towards
autocracy during the 1940s and, after World War II, moved rapidly into
the democratic basin, where it has remained with only minor fluctuations.
The United States shows a more gradual evolution towards the high-density
democratic cluster, with a reversal in recent years. Switzerland, by contrast,
begins and remains within the democratic basin throughout its entire history.

The diversity of these trajectories becomes clearer when quantifying
how much area of the regime space is explored over a time interval
$\Delta$. The time-averaged mean-squared displacement (TAMSD;
inset of Fig.~\ref{fig:fig1_trajectory}) reveals that individual
country TAMSDs vary over nearly two orders of magnitude. The wide
spread of individual TAMSD curves indicates weak ergodicity breaking
\cite{Bouchaud1992,BelBarkai2005,HeBurovMetzlerBarkai2008,Metzler2000,
SchulzBarkaiMetzler2014}, a hallmark of intermittent dynamics. In practical
terms, this means that countries do not ``sample" the political landscape
uniformly: some remain trapped in stable configurations for decades,
while others undergo rapid and repeated transitions. As a result, long-term
behaviour remains strongly country-specific rather than averaging out across
time.  Formally, in such systems, time averages along single trajectories
do not converge to the same value as averages taken across countries
\cite{Bouchaud1992,BelBarkai2005,HeBurovMetzlerBarkai2008,Metzler2000,
SchulzBarkaiMetzler2014}. Countries undergoing major shifts, such as Hungary,
display large displacements, whereas stable democracies, such as Switzerland,
show minimal displacement. At small lag times, individual TAMSDs scale close to
linearly with $\Delta$, though some appear slightly super- or sub-linear. At
large $\Delta$, fewer points contribute to the average, producing irregular
curves where country-specific trends dominate. For example, Japan's TAMSD
plateaus during its recent stability, while Hungary's decreases during
democratic backsliding.

\begin{figure}
\includegraphics[width=\columnwidth]{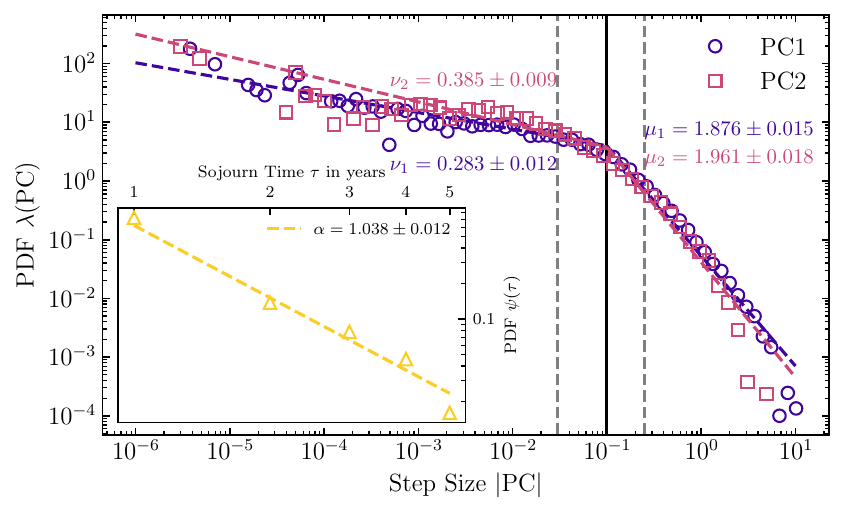}
\caption{\label{fig:hist}\textbf{Step size and sojourn time PDFs in regime
space.} Histograms show the PDFs of annual step sizes in PC1 and PC2;
inset shows the normalised sojourn time PDF. Dashed coloured lines indicate
maximum-likelihood power-law fits using a representative crossover point
$\mathrm{PC}_c=0.1 $ (black vertical line), with estimated exponents
reported in the panels. Grey dashed lines indicate the range of crossover
points considered.}
\end{figure}

\begin{figure}
\includegraphics[width=\columnwidth]{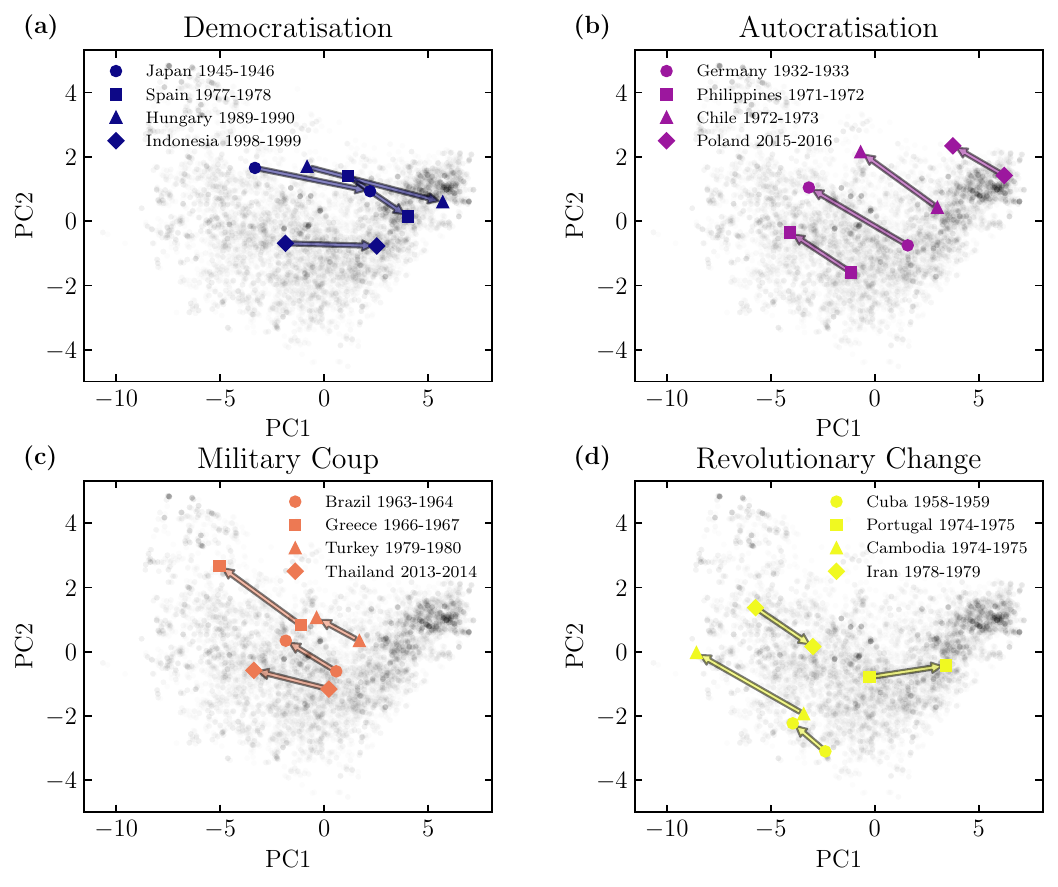}
\caption{\textbf{Representative examples of large regime transitions
in the PC1-PC2 political regime space.} Arrows indicate one-year
displacements associated with well-documented episodes of democratisation
(a), autocratisation (b), military coups (c), and revolutionary change
(d). These examples show that large movements in regime space correspond to
substantive institutional ruptures, providing qualitative grounding for the
heavy-tailed step-size statistics analysed in the main text.}
\label{fig:extreme_examples}
\end{figure}

A ``step" refers to the change in a country's position from one year to the
next in the PC1-PC2 space, and the step size distribution characterises how
large these changes typically are. Figure~\ref{fig:hist} shows the probability
density functions (PDFs) $\lambda_i(\mathrm{PC}i)$ $(i=1,2)$ of step sizes
on log-log axes. Both dimensions exhibit two distinct scaling regimes
separated by a crossover scale. Small steps are common and approximately
uniformly distributed, reflecting routine policy adjustments and incremental
institutional change. In contrast, large steps follow a heavy-tailed
PDF, corresponding to rare, abrupt regime shifts such as revolutions,
coups, or breakdowns of political order. Figure \ref{fig:extreme_examples}
presents representative examples of large one-year displacements associated
with well-documented institutional ruptures, including democratisation,
autocratisation, coups, and revolutions.  The crossover between these
regimes is sharper in PC1, while PC2 shows a smoother change in slope,
possibly reflecting a more continuous trade-off between civil liberties and
electoral capability.

We model the step size PDFs as bimodal power-laws with a crossover point
$\mathrm{PC}_c$ following
\begin{equation}
\label{eq:stepsize}
\lambda_i(\mathrm{PC}i)\propto
\begin{cases}
|\mathrm{PC}i|^{-\nu_i}, & \mathrm{PC}_l \le |\mathrm{PC}i| < \mathrm{PC}_c,\\
|\mathrm{PC}i|^{-\mu_i}, & |\mathrm{PC}i| \ge \mathrm{PC}_c,
\end{cases}
\end{equation}
where $\mathrm{PC}_l$ is the lower bound of the power-law (see Methods
\ref{Methods_exponent}). To characterise the tail behaviour, we estimate the
exponents $\mu_i$ for a range of crossover points $\mathrm{PC}_c\in[0.03,0.25]$
(grey dashed lines in Fig.~\ref{fig:hist}). Across this full range,
the maximum-likelihood estimates remain within $\mu\approx 1.6..2.4$,
i.e., near the critical regime $\mu=2$ where the mean step size diverges
(Appendix~\ref{appendixTransition}). This shows that the heavy-tailed
structure is robust and not an artefact of threshold choice. For clarity
in presentation, we therefore report exponent values using a representative
cutoff $\mathrm{PC}_c=0.1$. Together, these two scaling regimes indicate
that frequent small adjustments and rare large transformations characterise
how political systems change.

We next examine how long countries tend to remain in a given configuration
before such changes occur.

The sojourn time PDF $\psi(\tau)$ (inset of Fig.~\ref{fig:hist}) also appears
to follow a power-law,
\begin{equation}
\psi(\tau)=\alpha\tau_0^\alpha\tau^{-(1+\alpha)},
\end{equation}
with a lower cutoff $\tau_0=1$ year, due to the temporal resolution.
Here, sojourn times correspond to periods during which a country remains
in approximately the same institutional configuration. A heavy-tailed PDF,
therefore, implies that while many regimes change quickly, others persist
for unusually long periods, yielding extended stability punctuated
by rare, abrupt shifts. The discrete maximum-likelihood estimator
(Eq.~(\ref{eq:discrete_mle})) yields the exponent $\alpha\approx1$. Thus,
sojourn times are likewise near a critical threshold where mean values diverge,
allowing both short and long intervals of political stasis.  Together, these
results reveal that countries exhibit near-critical, weakly non-ergodic regime
dynamics, characterised by heavy-tailed step sizes and sojourn times.

\subsection{Temporal stability of the regime space}

\begin{figure}
\includegraphics[width=\columnwidth]{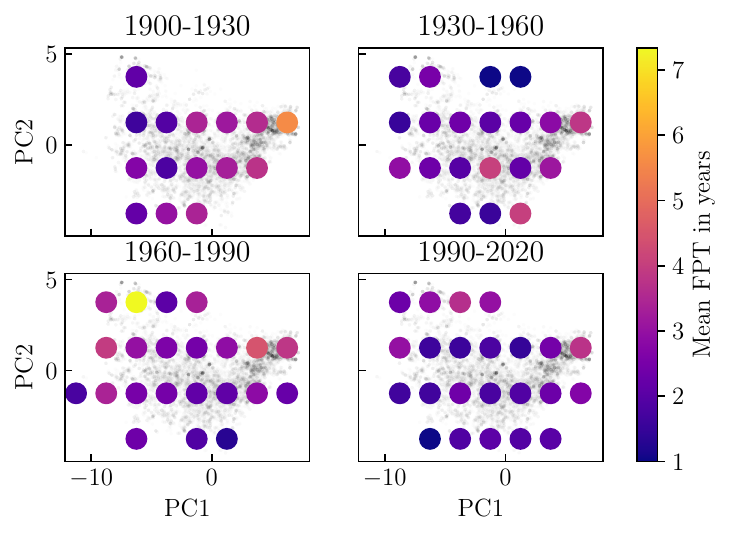}
\caption{\label{fig:fpt} 
\textbf{Mean first-passage times (FPTs) in the regime space for successive
30-year windows.} Warm colours indicate longer FPTs and thus greater local
political stability, corresponding to prolonged residence near a given
institutional configuration.} \end{figure}

Having characterised how countries move through the regime space, we now
address a natural question: are there regions in this political landscape where
regimes persistently stabilise, or is stability itself historically contingent?
To identify regions of local stability, we compute first-passage times (FPTs)
\cite{Redner2001,Metzler2014}, defined as the time a trajectory remains within
a neighbourhood before exiting (see Methods \ref{sec:FPT}). This measure
captures how long political systems persist near a given institutional
configuration before undergoing substantial change. Spatially averaging
FPTs by starting position reveals heterogeneous stability basins that shift
over time (Fig.~\ref{fig:fpt}).

In the early period (1900-1930), two main zones of extended stability are
visible. One lies around $\mathrm{PC}1\approx0$, encompassing colonies such
as South Africa and monarchies like Hungary or Japan. The other corresponds
to the democratic basin at high PC1 and low PC2 values.

Between 1930 and 1960, the hybrid regime region becomes increasingly isolated,
with its surrounding area showing markedly reduced stability. This transition
coincides with the erosion of imperial and monarchical systems worldwide. At
the same time, the democratic area retains reduced but still moderately high
FPTs, while the autocratic extreme begins to exhibit slightly longer FPTs,
suggesting the emergence of a new stable configuration at this pole of the
regime space.

From 1960 to 1990, the centres of stability shift decisively towards these
extremes. Both the high PC1 democratic region and the low PC1 autocratic
region display elevated FPTs. In the case of the autocratic region, up to
twice the values seen in earlier decades appear. Deeply autocratic regimes
such as Albania's remain virtually static during this period, contributing
to the extended persistence observed in that region.

In the most recent period (1990-2020), the democratic and autocratic poles
remain the most stable regions, with reduced but similarly long FPTs in both
areas. The FPT in the surrounding areas decays slowly to very short times
$\approx1$ year, indicating broader but shallower stability zones.

Across the full time span, the democratic basin persists as the only consistent
stable region. The early hybrid regime zone dissolves mid-century, giving
way to an autocratic basin that weakens towards the present. Overall, these
results reveal a rugged, shifting landscape of political stability, in which
enduring regime configurations are rare and increasingly polarised between
democratic and autocratic extremes. Next, we test whether the dynamics can
arise from a minimal stochastic mechanism.

\subsection{Modelling regime dynamics}

\begin{figure}
\includegraphics[width=\columnwidth]{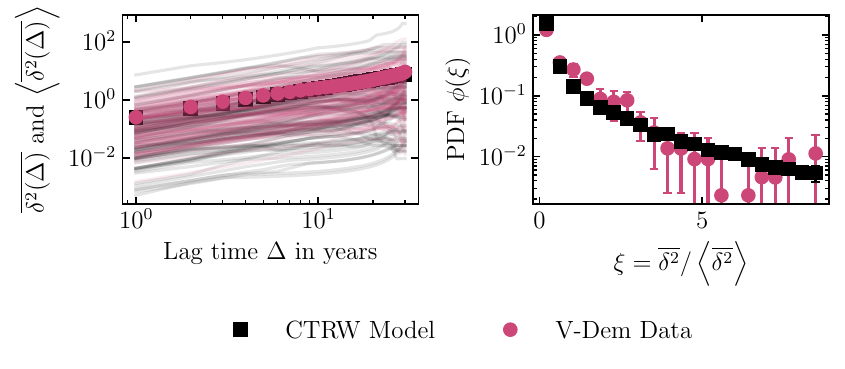}
\caption{\label{fig:datavmodell} \textbf{Time-averaged mean-squared
displacement (TAMSD) for empirical trajectories and continuous time random walk
(CTRW) simulations.} Lines show individual TAMSDs, symbols indicate their
ensemble averages, and the corresponding amplitude scatter PDF $\phi(\xi)$
is shown alongside.}
\end{figure}

The combination of heavy-tailed step sizes, scale-free sojourn times, and weak
ergodicity breaking is characteristic of systems well described by a CTRW,
making it a natural minimal model for regime dynamics. In a CTRW, a walker makes
jumps separated by random sojourn times \cite{MontrollWeiss1965,Metzler2000},
which here correspond to regime shifts and periods of institutional inertia.
We simulate a two-dimensional CTRW parameterised by the empirically estimated
step size and sojourn time PDFs.

We compare the simulated trajectories to empirical regime paths from 1990-2020,
where uninterrupted data are available for $N=108$ countries. As shown in
Fig.~\ref{fig:datavmodell}, the ensemble-averaged TAMSD $\langle\overline{
\delta^2} \rangle$ from the CTRW closely matches the empirical behaviour.
Moreover, the amplitude scatter of individual TAMSDs, quantified by the PDF
$\phi(\xi)$ \cite{HeBurovMetzlerBarkai2008,MetzlerJeonCherstvyBarkai2014}, is
reproduced as well, capturing the characteristic spread associated with weak
ergodicity breaking. Similar levels of agreement hold across the full robust
range of PC$_c$ values (Appendix~\ref{appendixTransition}). Thus, the model
captures both the mean dynamical behaviour and the trajectory-to-trajectory
variability. The overall agreement is striking for such a simple model.

These results demonstrate that near-critical CTRW dynamics are sufficient to
reproduce the essential statistical structure of regime evolution, including
intermittent movement, heterogeneous stability, and persistent non-ergodicity.

\section{\label{sec:discussion}Discussion}

Political regime evolution is often framed either as universal development
\cite{Lipset1959,North1981,Fukuyama1989} or as the result of historical
events and path dependence \cite{Pierson2000,Mahoney2000}. Our findings
show that both views are correct, but incomplete: regime trajectories are
historically specific, yet they also follow shared dynamical patterns.
Countries move through political regime space in a manner characterised by
intermittent dynamics, combining periods of stability with occasional
large-scale transitions.  The PDFs of both step sizes and sojourn
times follow heavy-tailed, near scale-free forms, and the TAMSD
exhibits strong trajectory-to-trajectory variability.  Together,
these signatures indicate weak ergodicity breaking, in which long-term
political trajectories do not converge towards a shared path but remain
strongly shaped by historical contingencies and institutional legacies
\cite{Bouchaud1992,BelBarkai2005,HeBurovMetzlerBarkai2008,Metzler2000,
SchulzBarkaiMetzler2014}.

The stability landscape inferred from first-passage times shows that
persistent regime configurations are sparse and historically shifting.
A basin associated with hybrid systems is evident in the early 20th century
but dissolves mid-century.  A stability region appears at the autocratic
extreme during the mid-century period, before weakening in recent decades.
In contrast, the democratic basin is the only stability region that persists
across the entire time span.  These patterns indicate a rugged and evolving
landscape of viability, in which the stability of political configurations
depends on the broader historical context.

When stable states are sparse, movement patterns with tail exponents near $\mu
\approx 2$ maximise the likelihood of reaching them \cite{Viswanathan1996,
Viswanathan1999,Viswanathan2011,Lomholt2005,Lomholt2008,Palyulin2019}. The
same exponents and the resulting interplay between long residence times and
occasional large transitions arise in heterogeneous environments in which
walkers must balance the risks and rewards of exploration \cite{Boyer2006}.
The empirical step size and sojourn time PDFs fall in this range, indicating
that countries move as if navigating a heterogeneous landscape of viable
configurations. Concurrently, power-law distributions of sojourn times are
characteristic of diffusion and transport dynamics in living biological cells
\cite{Song2018,Weigel2011} as well as in human travel \cite{Brockmann2006}.
The implication is not that countries choose such movement strategies,
but that they arise from the structure of the stability landscape.

A CTRW parameterised by empirically measured sojourn time and step size PDFs
reproduces both the average regime movement and the degree of weak ergodicity
breaking. This suggests that the interplay between long residence times and
rare, large transitions is sufficient to generate the observed patterns.
Political systems, therefore, appear to operate near a critical boundary
between rigidity and volatility: stable enough to preserve institutions in the
absence of shocks, yet sufficiently flexible to undergo abrupt reconfiguration
when internal or external pressures accumulate.

The model abstracts away from the causes of specific transitions, such as
economic crises, social movements, leadership turnovers or international
alignment. The examples shown in Fig.~\ref{fig:extreme_examples} correspond
to well-documented episodes of democratisation, autocratisation, military
coups, and revolutionary change, including Japan's post-war transition,
Spain and Hungary's democratic openings, Germany's 1933 authoritarian
consolidation, Chile's 1973 coup, and the Iranian and Cuban revolutions.
Although historically distinct, these events contribute to the same
heavy-tailed distribution of step sizes, indicating that diverse political
ruptures share common stochastic dynamics. Future work could integrate
external influences like geopolitical shocks, economic crises or collective
movements by including additional stochastic or deterministic forces that
act on the countries. However, verifying such extensions is challenging given
the annual resolution and fragmented nature of the empirical trajectories.

Taken together, these results show that political regime change is shaped
both by country-specific histories and by general dynamical patterns. While
the events driving transitions may differ across cases, the way regimes
move through the political space follows a shared structure. These movements
reflect the shape of the regime landscape and the limited number of stable
configurations within it.

\section{\label{sec:methods}Methods}

\subsection{Data preprocessing} 

To ensure continuity when computing time-averaged observables, we restrict
analysis to trajectory segments containing at least ten consecutive years
without missing entries.  For the comparison between empirical data and model
simulations, we use the period 1990-2020, during which 108 uninterrupted
country trajectories are available.

\subsection{Exponent estimation}
\label{Methods_exponent}

For visualisation, we plot histograms of the step size PDFs in
Fig.~\ref{fig:hist}, where bins are chosen on a log scale using the outlier
resilient Freedman Diaconis estimator \cite{FreedmanDiaconis1981}. To estimate
the exponents $\nu_i$ and $\mu_i$ of the PDFs $\lambda_i(\mathrm{PC}i)$
(Eq.~(\ref{eq:stepsize})) we use maximum likelihood estimation (MLE). The
lower bound of the PDF PC$_l=10^{-6}$, required for exponent estimation, is
chosen as the order of magnitude of the smallest observed step size. For the
tail exponent $\mu_i$ the exponent by MLE is given by \cite{Clauset2009}
\begin{equation}
\label{eq:continouous_mle}
\mu_i=1+n\left[\sum_{j=1}^n\ln\left(\frac{|\mathrm{PC}i|_j}{\mathrm{PC}_c}\right)
\right]^{-1}.
\end{equation}
For the bounded first regime we determine $\nu_i$ numerically by maximising the
log-likelihood \cite{Langlois2014}
\begin{equation}
-\nu\sum_{j=1}^m\ln|\mathrm{PC}i|_j+m\ln\left(\frac{1-\nu}{(\mathrm{PC}_c)^{1
-\nu}-(\mathrm{PC}_l)^{1-\nu}}\right).
\end{equation}

For the discrete sojourn times, no binning is needed, and the normalised counts
are displayed in the inset of Fig.~\ref{fig:hist}. We use the corresponding
discrete MLE estimator \cite{Clauset2009}
\begin{equation}
\label{eq:discrete_mle}
\alpha=N\left[\sum_{i=1}^N\ln\left(\frac{\tau_i}{\tau_0-1/2}\right)\right]^{-1}.
\end{equation}

\subsection{First passage time}
\label{sec:FPT}

To quantify local stability in the regime space, we compute the FPT for
each trajectory. For a given starting point, the FPT is defined as the
first time step at which the trajectory exits a circle of radius $ 0.1$
in the PC1-PC2 plane. This radius corresponds to the approximate crossover
scale between small and large regime changes (see Fig.~\ref{fig:hist} and
Appendix~\ref{appendixTransition}). If a trajectory does not cross the
threshold within the observable period, the FPT is treated as undefined and
excluded from averaging. FPTs are computed for every possible starting
year along each trajectory and then spatially binned across the PC1-PC2
plane using square bins of side length $2.5$. This bin size ensures that
each bin contains data from multiple countries while avoiding unnecessary
coarse-graining. The mean FPT per bin defines the stability landscape shown
in Fig.~\ref{fig:fpt}.

\subsection{Weak ergodicity breaking}

The TAMSD of a trajectory $x(t)$ of length $T$ is defined as
\cite{MetzlerJeonCherstvyBarkai2014}
\begin{equation}
\label{eq:TAMSD}
\overline{\delta^2(\Delta,T)}=\frac{1}{T-\Delta}\int_0^{T-\Delta}[x(t+\Delta)
-x(t)]^2\,\mathrm{d}t,
\end{equation}
where $\Delta$ is the lag time. The TAMSD quantifies how much the trajectory
spreads on average over the regime space over a given time interval. We use
the TAMSD instead of the ensemble-averaged MSD because country trajectories
have different lengths and starting conditions, making the construction of
an ensemble difficult.

For ergodic processes such as Brownian motion \cite{Einstein1905,Langevin1908,
MetzlerJeonCherstvyBarkai2014}, the time and ensemble averages coincide in the
long-time limit, and individual TAMSD curves collapse onto the ensemble mean.
In contrast, many complex systems exhibit a broad spread of TAMSD values across
trajectories. The averaged TAMSD,
\begin{equation}
\left\langle\overline{\delta^2(\Delta)}\right\rangle=\frac{1}{N}\sum_{i=1}^N
\overline{\delta^2(\Delta)}_i,
\end{equation}
is then not representative of any individual realisation. This phenomenon is
known as weak ergodicity breaking \cite{Bouchaud1992,BelBarkai2005,
HeBurovMetzlerBarkai2008,Metzler2000,SchulzBarkaiMetzler2014}.

A CTRW naturally produces weak ergodicity breaking when either the mean sojourn
time \cite{BelBarkai2005,HeBurovMetzlerBarkai2008,SchulzBarkaiMetzler2014} or
the second moment of the step size PDF diverges \cite{Vahabi2013,Froemberg2013}.
In such cases, the corresponding diffusion equation becomes fractional in both
space and time. In particular, when the mean sojourn time diverges, the process
exhibits memory, meaning the probability of being at a particular position
depends on the full history of the trajectory, rather than only its current
state.

\subsection{CTRW simulations}

The V-Dem indices are bounded by construction \cite{VDem}, and therefore their
principal components are also bounded. To ensure that simulated trajectories
remain realistic, we impose an upper cutoff on the step size PDFs. We choose
these upper bounds as the largest observed empirical step sizes in each
dimension ($\mathrm{PC1}_{\max}\approx11.16$ and $\mathrm{PC2}_{\max}\approx
5.53$), which prevents unrealistically large jumps while avoiding additional
parameterisation of the spatial bounds.

The CTRW is simulated as follows. At each update, a
sojourn time $\tau$ is drawn from the empirical sojourn time PDF $\psi(\tau)$
and rounded to the nearest integer (reflecting the yearly temporal resolution).
After the sojourn period, step sizes in PC1 and PC2 are independently sampled
from their respective heavy-tailed PDFs, subject to the upper bounds above.
This process is repeated until the desired total trajectory length is reached.

To compare to the empirical period 1990-2020, we simulate an ensemble of $10^4$
trajectories, each with a duration of $T=31$ years.

\appendix
\renewcommand\thefigure{\thesection.\arabic{figure}} 
\setcounter{figure}{0}

\section{Robustness of the crossover point}\label{appendixTransition}

\begin{figure}
\includegraphics[width=\columnwidth]{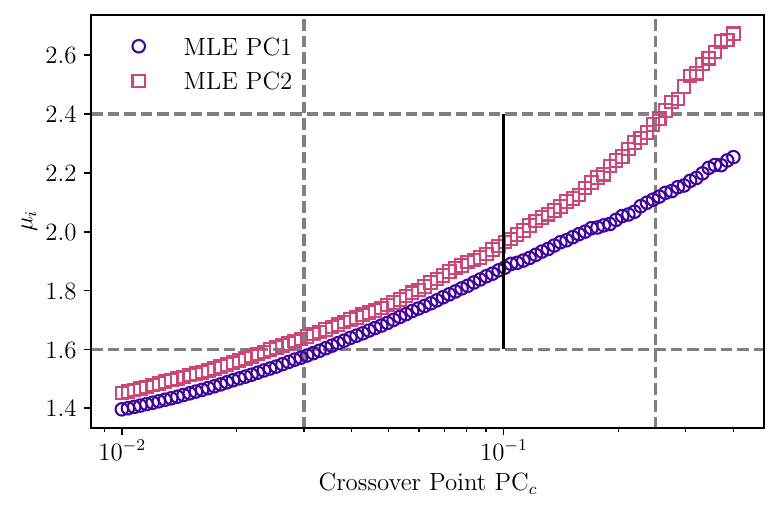}
\caption{\label{fig:likelihood_cutoffs} Maximum-likelihood estimates of the
tail exponents $\mu_i$ of the step size PDFs $\lambda_i(\mathrm{PC}i)$
$(i=1,2)$ for different crossover points $\mathrm{PC}_c$. The vertical grey
lines indicate the range of reasonable crossover points $\mathrm{PC}_c\in[0.03,
0.25]$, and the black line marks the representative value $\mathrm{PC}_c=0.1$
used in the main text.}
\end{figure}

The exponent of the power-law tail depends on the choice of the crossover
point $\mathrm{PC}_c$, so we assess the influence of this choice here.
Figure~\ref{fig:likelihood_cutoffs} shows the estimated tail exponents over
a wide range of $\mathrm{PC}_c$ values. Within the empirically reasonable
range $\mathrm{PC}_c\in[0.03,0.25]$, the tail exponents remain within $\mu_i
\approx 1.6..2.4$, i.e., close to the critical regime $\mu=2$. The absence of
a plateau is consistent with the bounded nature of the V-Dem data, suggesting
a PDF with a cutoff, potentially exponential in nature.

\begin{figure}
\includegraphics[width=\columnwidth]{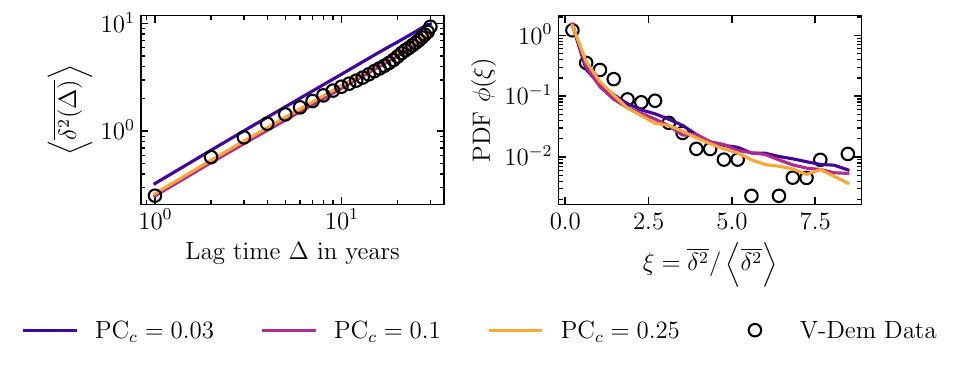}
\caption{Ensemble-averaged TAMSD $\left\langle\overline{\delta^2(\Delta)}
\right\rangle$ and TAMSD amplitude scatter PDFs $\phi(\xi)$ for CTRW
simulations with different crossover points, compared to empirical data
(1990-2020).}
\label{fig:dynamics_cutoffs}
\end{figure}

To evaluate the effect of this variation on the dynamics, we simulate
the CTRW using the exponent values corresponding to the lower bound,
upper bound, and representative crossover point $\mathrm{PC}_c=0.1$.
Figure~\ref{fig:dynamics_cutoffs} shows that both the ensemble-averaged
TAMSD $\langle\overline{\delta^2(\Delta)}\rangle$ and the amplitude scatter
PDF $\phi(\xi)$ remain in close agreement with the empirical data in all cases.

Thus, the heavy-tailed structure of regime shifts and the resulting CTRW
dynamics are robust to the choice of $\mathrm{PC}_c$, and do not depend on
fine-tuning of this parameter. This confirms that the key results reflect
structural features of the empirical PDFs rather than the specific choice
of crossover threshold.

\section*{Data availability}

The V-Dem Dataset is publicly available at \url{https://v-dem.net/data/},
and the data subset used in this work can be found at
\url{https://github.com/JoshuaUhlig/RegimeDynamics}.

\section*{Code availability}

All code used in this study is available at
\url{https://github.com/JoshuaUhlig/RegimeDynamics}.

\bibliography{bibliography}

\begin{acknowledgments}

We thank Erik Kalz and Andrey G. Cherstvy for helpful discussions. We
acknowledge the German Science Foundation (DFG, Grant ME 1535/22-1 and CRC
Data assimilation, Grant 318763901, project B10) for support.

\end{acknowledgments}

\end{document}